\documentclass[aps,prd,twocolumn,groupedaddress]{revtex4-2}

\usepackage{amsmath,amssymb,amsfonts,amsthm,bm,braket,cases,mathrsfs,subfigure}
\usepackage{graphicx}
\renewcommand{\vec}[1]{\bm{#1}} 
\newcommand{\mat}[1]{\mathbf{\bm{#1}}}   
\newcommand{\flavor}[1]{\mathsf{#1}}    
\newcommand{\tr}{\operatorname{tr}}
\newcommand{\re}{\operatorname{Re}}
\newcommand{\im}{\operatorname{Im}}
\newcommand{\diag}{\operatorname{diag}}

\newtheorem{lem}{Lemma}

\begin{document}

\title{Fast neutrino flavor instability and neutrino flavor lepton number crossings}

\author{Taiki Morinaga}
\affiliation{Graduate School of Advanced Science and Engineering, Waseda University, 3-4-1 Okubo, Shinjuku, Tokyo 169-8555, Japan}

\date{\today}

\begin{abstract}
In this paper, we show the equivalency between the existence of fast neutrino flavor instability and that of neutrino flavor lepton number (NFLN) crossings, which indicates that an NFLN angular distribution takes both signs.
The veracity of this proposition has been uncertain and sometimes controversial despite its indispensability in the flavor evolutions of dense neutrinos.
This study clarifies that the occurrence of an NFLN crossing is both necessary and sufficient for fast instability.
\end{abstract}

\pacs{}

\maketitle

\section{Introduction}
In an environment such as a supernova where a large number of neutrinos are present, neutrino oscillations exhibit nonlinear behaviors due to the self-interaction of neutrinos~\cite{Sawyer2005,Duan2006,Hannestad2006,Raffelt2007,Raffelt2007a,Dasgupta2008,Dasgupta2009,Duan2010,Patwardhan2019,Rrapaj2019,Rrapaj2020}.
It is quite difficult to solve the kinetic equations that describe this phenomenon, called collective neutrino oscillations, because of the enormous computational cost;
the spatial and temporal scales of the oscillations are usually much smaller than those of a supernova, and very fine grids are needed in the momentum space to obtain even qualitatively correct behaviors~\cite{Sarikas2012}.

However, collective neutrino oscillations do not always occur by working against the matter suppression~\cite{Wolfenstein1978,Wolfenstein1979} of flavor conversions when dense matter exists.
The conditions crucial for the occurrence of collective neutrino oscillations have been investigated by linear stability analysis~\cite{Sawyer2009,Banerjee2011,Mirizzi2012,Mirizzi2012a,Mirizzi2013,Chakraborty2014,Chakraborty2014a,Abbar2015,Dasgupta2015,Sawyer2016,Chakraborty2016,Dasgupta2017,Izaguirre2017,Airen2018,Chakraborty2020}.
In particular, fast flavor instability~\cite{Sawyer2009,Chakraborty2016,Dasgupta2017,Izaguirre2017,Abbar2018,Airen2018,Dasgupta2018,Yi2019}, which is a kind of unstable mode and whose spatial and temporal scales are proportional to the inverse of the density of neutrinos, has attracted much attention~\cite{Capozzi2017,Dasgupta2018b,Abbar2018b,Martin2019,DelfanAzari2019,Abbar2020a,Johns2020,Abbar2020c,Bhattacharyya2020,Shalgar,Bhattacharyya2021,Martin2021}.
Indeed, some studies have discussed the possibilities of fast flavor conversion in various regions, such as the regions inside~\cite{DelfanAzari2020,Glas2020} and just above~\cite{Capozzi2019a} a protoneutron star and the preshock region~\cite{Morinaga2020a} in a supernova.
Additionally, asymmetric neutrino emissions~\cite{Abbar2018a,Nagakura2019,Abbar2020a,Abbar2020c} and breaking the degeneracy of heavy leptonic neutrinos~\cite{Chakraborty2020,Capozzi2020,Capozzi2020b} can affect the possible regions for fast flavor conversion.

It is important that all of these studies focus on crossings of the neutrino flavor lepton number (NFLN) angular distributions.
Many studies have suggested that a fast instability appears when the difference between the NFLN angular distributions of 2 flavors crosses with 0.
However, whether an NFLN crossing is necessary and/or sufficient is not known.
In particular, the veracity of its sufficiency is sometimes dubious and even controversial.
For example, Ref.~\cite{Johns2020a} concluded that the presence of an NFLN crossing is not sufficient for fast instability under the assumption of axisymmetry and spatial homogeneity.
According to Ref.~\cite{Capozzi2020}, a ``shallow crossing" in the electron lepton number distribution does not generate instability.
However, these results occur because artificially imposed symmetries hinder the development of unstable modes.

In this paper, we show that the existence of fast instability is equivalent to that of NFLN crossings.
Mathematical proof of this proposition has been a missing link in the study of fast flavor conversion.
In addition, we find that spurious instability~\cite{Sarikas2012} by the discretization of spectra does not appear over time, unlike stationary solutions.
If an NFLN crossing exists, at least modes with the wave vector $\vec{k}$ around the ``crossing direction," at which the NFLN angular distributions of 2 flavors cross each other, exhibit instability.
This study clarifies the condition for fast neutrino flavor instability and plays a crucial role in the elucidation of collective neutrino oscillations.

\section{Fast neutrino flavor instability}
\subsection{Kinetic equation}
We consider the density matrix of $N_\mathrm{f}$-flavor neutrinos (antineutrinos) $\flavor{f}$ ($\bar{\flavor{f}}$), which is $N_\mathrm{f} \times N_\mathrm{f}$ matrices depending on the spacetime position $x$, energy $E$ and flight direction $\vec{v}$.
Through the introduction of the density matrix with negative energy $-E < 0$ as $\flavor{f}(-E) \equiv -\bar{\flavor{f}}(E)$, their evolutions are described collectively by the kinetic equation~\cite{Sigl1993,Yamada2000,Cardall2008,Vlasenko,Kartavtsev2020}
\begin{align}
    v\cdot\partial\flavor{f}(x,\Gamma) = -i[\flavor{H}(x,\Gamma),\flavor{f}(x,\Gamma)],
    \label{eq:Kinetic}
\end{align}
where $\Gamma \equiv (E, \vec{v})$ and $(v^\mu) \equiv (1, \vec{v})$, and the Hamiltonian $\flavor{H}$ is expressed as
\begin{align}
    \flavor{H}(x,\Gamma) \equiv \dfrac{\flavor{M}^{2}}{2E} + v\cdot\flavor{J}(x).
\end{align}
The first term of $\flavor{H}$ is the vacuum mixing term, which mixes the neutrino flavors by the off-diagonal components of the mass square matrix $\flavor{M}^2$.
The second term reflects the forward scattering of neutrinos on leptons with
\begin{align}
    \flavor{J}^\mu(x)\equiv \sqrt{2}G_F\left[\diag\left(\{j^\mu_\alpha(x)\}\right) + \int d\Gamma \flavor{f}(x,\Gamma)v^\mu\right],
\end{align}
where $j_\alpha$ is the lepton number current of charged leptons $\alpha$ and $\int d\Gamma \equiv (2\pi)^{-3}\int_{-\infty}^\infty dE E^2 \int d\vec{v}$.

\subsection{Dispersion relation of the fast mode}
Fast neutrino flavor instability is the instability of the flavor eigenstates $\flavor{f} = \diag(\{f_{\nu_\alpha}\})$ when the vacuum mixing term is neglected because it is minor compared to the self-interactions of neutrinos~\cite{Izaguirre2017,Airen2018}.
To consider this class of instability, we omit $\flavor{M}^2$ and linearize Eq.~(\ref{eq:Kinetic}) as
\begin{align}
    &v\cdot \left\{i\partial - J_{0\alpha}(x) + J_{0\beta}(x)\right\}S_{\alpha\beta}(x,\Gamma)\nonumber\\
    &+ \left(f_{\nu_\alpha}(x, \Gamma) - f_{\nu_\beta}(x, \Gamma)\right)\sqrt{2}G_F\int d\Gamma' v\cdot v'S_{\alpha\beta}(x,\Gamma') = 0
    \label{eq:KineticLinearized}
\end{align}
for the Hermitian matrix $\flavor{S}$ given as $\flavor{f} = \diag(\{f_{\nu_\alpha}\}) + \flavor{S}$, where $J_{0\alpha}^\mu(x) \equiv \sqrt{2}G_F\left[j^\mu_\alpha(x) + \int d\Gamma f_{\nu_\alpha}(x,\Gamma)v^\mu\right]$.

We neglect the spatial and temporal variations in $\{j_\alpha\}$ and $\{f_{\nu_\alpha}\}$ and substitute the plane wave ansatz $\flavor{S}(x, \Gamma) = \tilde{\flavor{S}}(k, \Gamma)e^{-ik\cdot x}$ into Eq.~(\ref{eq:KineticLinearized}) to derive the dispersion relation (DR), where $(k^\mu) \equiv (\omega, \vec{k})$ denotes an angular frequency $\omega$ and wave vector $\vec{k}$.
The diagonal components of Eq.~(\ref{eq:KineticLinearized}) yield DR $v\cdot k = 0$, which does not generate instabilities, and the others do
\begin{align}
    \Delta_{\alpha\beta}(k) \equiv \det\mat{\Pi}_{\alpha\beta}(k) = 0,
    \label{eq:DR}
\end{align}
where
\begin{align}
    \Pi^{\mu\nu}_{\alpha\beta}(k) \equiv \eta^{\mu\nu} + \int \frac{d\vec{v}}{4\pi} G_{\alpha\beta}(\vec{v})\dfrac{v^\mu v^\nu}{v\cdot (k - J_{0\alpha} + J_{0\beta})}.
    \label{eq:PolarizationTensor}
\end{align}
$G_{\alpha\beta}(\vec{v}) \equiv \sqrt{2}G_F\int_{-\infty}^\infty \frac{dE E^2}{2\pi^2}\left(f_{\nu_\alpha}(\Gamma) - f_{\nu_\beta}(\Gamma)\right)$ is the difference between the NFLN angular distribution for $\nu_\alpha$ and $\nu_\beta$.
We note that $\Delta_{\alpha\beta}(k) = \Delta_{\beta\alpha}(-k)$ is satisfied and that all the $[N_\mathrm{f}(N_\mathrm{f} - 1)/2]$-independent equations of Eq.~(\ref{eq:DR}) are candidates for instabilities.
In the following discussions, we consider one of them and omit the indices denoting flavor.
Additionally, we set $J_0 = 0$ because $J_0$ shifts only the real parts of the wave vector $k$ and does not affect the instability.
In addition, we assume that $G$ is continuous, which is a natural assumption for treating realistic systems.

Because $\Delta(\omega, \vec{k}) = \overline{\Delta(\overline{\omega}, \vec{k})}$ is satisfied for $\vec{k} \in \mathbb{R}^3$, the complex conjugate pair $(\omega, \vec{k})$ and $(\overline\omega, \vec{k})$ are both solutions of Eq.~(\ref{eq:DR}).
Therefore, if there exist nonreal $\omega$ values for $\vec{k} \in \mathbb{R}^3$, $\flavor{S}$ grows exponentially.
Note that a nonreal $\vec{k}$, which is sometimes called a ``spatial instability'', does not directly play a role in spatiotemporal evolutions and does not even guarantee the spatial growth of perturbations imposed ceaselessly at some spatial point~\cite{Sturrock1958,Briggs1964,Landau1997,Capozzi2017,Yi2019,Morinaga2020,Morinaga2021}.

\section{Equivalency between fast instability and NFLN crossings}
In this section, we show that the necessary and sufficient condition for the existence of fast instability is that of NFLN crossings.

\subsection{Necessary condition}
First, we focus on the necessary condition:
if there exist $\omega \notin \mathbb{R}$ and $\vec{k} \in \mathbb{R}^3$ such that $\Delta(k) = 0$, $G(\vec{v})$ takes both positive and negative values.

We define
\begin{align}
    \sigma \equiv& \im \omega\\
    (\kappa^\mu) \equiv& (\re\omega, \vec{k})
\end{align}
to separate the real and imaginary parts of $\omega$.
Then, $\mat{\Pi}$ can be decomposed as
\begin{align}
    \Pi^{\mu\nu}(k) = R^{\mu\nu}(k) - i I^{\mu\nu}(k),
\end{align}
where we define
\begin{align}
    R^{\mu\nu}(k) \equiv& \eta^{\mu\nu} + \int \frac{d\vec{v}}{4\pi} G(\vec{v})\dfrac{v^\mu v^\nu v\cdot \kappa}{(v\cdot \kappa)^2 + \sigma^2}\\
    I^{\mu\nu}(k) \equiv& \sigma\int \frac{d\vec{v}}{4\pi} G(\vec{v})\dfrac{v^\mu v^\nu}{(v\cdot \kappa)^2 + \sigma^2}.
\end{align}
The symmetric tensor $\mat{I}$ can be diagonalized by an orthogonal matrix $\mat{V} \in O(4, \mathbb{R})$ as
\begin{align}
    V^\mu{}_\sigma V^\nu{}_\rho I^{\sigma\rho} = D^{\mu\nu},
\end{align}
where $\mat{D}$ is a real diagonal matrix whose $(\mu, \mu)$-component is given as
\begin{align}
    D^{\mu\mu}(k) = \sigma \int \frac{d\vec{v}}{4\pi} G(\vec{v})\dfrac{\left(V^\mu{}_\nu(k) v^\nu\right)^2}{(v\cdot \kappa)^2 + \sigma^2}.
    \label{eq:Diagonal}
\end{align}
Then, $\mat{\Pi}$ can be expressed as
\begin{align}
    \Pi^{\mu\nu} = \left(V^{-1}\right)^\mu{}_\sigma \left(V^{-1}\right)^\nu{}_\rho \left(\tilde{R}^{\sigma\rho} - iD^{\sigma\rho}\right)
\end{align}
with $\tilde{R}^{\mu\nu} \equiv V^\mu{}_\sigma V^\nu{}_\rho R^{\sigma\rho}$, and Eq.~(\ref{eq:DR}) is equivalent to $\det\left(\mat{\tilde{R}}(k) - i \mat{D}(k)\right) = 0$, which means that there exists a nontrivial 4-vector $a$ such that $\tilde{R}^{\mu\nu} a_\nu = iD^{\mu\nu} a_\nu$.
From this equation, we can obtain $\overline{a}_\mu \tilde{R}^{\mu\nu} a_\nu = i\overline{a}_\mu D^{\mu\nu} a_\nu$ and $\overline{a}_\mu \tilde{R}^{\mu\nu} a_\nu = -i\overline{a}_\mu D^{\mu\nu} a_\nu$, whose difference yields
\begin{align}
\sum_\mu D^{\mu\mu}|a_\mu|^2 = 0.
\label{eq:ConditionForDR}
\end{align}
If $G(\vec{v})$ does not change its sign for all $\vec{v}$ and $\sigma \neq 0$, all the diagonal components of $\mat{D}$ have the same sign as $\sigma G$ from Eq.~(\ref{eq:Diagonal}) and cannot satisfy Eq.~(\ref{eq:ConditionForDR}).
Therefore, $G$ must take both positive and negative values for $\sigma$ to be nonzero.

We notice that the above discussion is valid even if $G$ is discretized as $G(\vec{v}) = \sum_i G_i \delta(\vec{v} - \vec{v}_i)$. If we consider stationary solutions, the discretization of spectra sometimes suffers from spurious instabilities~\cite{Sarikas2012}. On the other hand, when we solve time evolutions, spurious instability does not appear by discretization.

\subsection{Sufficient condition}
The remaining task is to show the sufficient condition:
if $G(\vec{v})$ takes both positive and negative values, there exist $\omega \notin \mathbb{R}$ and $\vec{k} \in \mathbb{R}^3$ such that $\Delta(k) = 0$.

By introducing
\begin{align}
    n^\mu \equiv \frac{k^\mu}{\omega},
\end{align}
$\mat{\Pi}$ can be expressed as
\begin{align}
    \Pi^{\mu\nu}(k) = \eta^{\mu\nu} + \frac{1}{\omega}T^{\mu\nu}(\vec{n}),
\end{align}
where
\begin{align}
    T^{\mu\nu}(\vec{n}) \equiv \int \frac{d\vec{v}}{4\pi} G(\vec{v})\dfrac{v^\mu v^\nu}{v\cdot n},
\end{align}
whose integral converges for $\vec{n}$ in the open unit ball $B \equiv \{\vec{n}\in\mathbb{R}^3||\vec{n}| < 1\}$.
Here, $\tr \mat{T} = 0$ due to $v^\mu v_\mu = 0$, where the trace of the natural powers of a tensor $\mat{A}$ is defined as $\tr \mat{A}^m \equiv A^{\mu_1}{}_{\mu_2} A^{\mu_2}{}_{\mu_3}\cdots A^{\mu_m}{}_{\mu_1}$.
Then, the DR is the zeros of the quartic function of $\omega$ (see Appendix~A):
\begin{align}
    \tilde\Delta(\omega, \vec{n}) \equiv& -\omega^4\Delta(k) = \det\left(\omega\delta^\mu_\nu + T^\mu{}_\nu(\vec{n})\right)\nonumber\\
    =& \omega^4 - \frac{1}{2}\tr \mat{T}^2(\vec{n})\omega^2 + \frac{1}{3}\tr \mat{T}^3(\vec{n})\omega\nonumber\\
    &+ \frac{1}{8}\left(\tr \mat{T}^2(\vec{n})\right)^2 - \frac{1}{4} \tr \mat{T}^4(\vec{n}).
    \label{eq:Quartic}
\end{align}
Henceforth, we express a solution for $\omega$ of $\Delta(k) = 0$ as $\omega(\vec{k})$ and that of $\tilde\Delta(\omega, \vec{n}) = 0$ as $\omega(\vec{n})$.
It should be noted that $\omega(\vec{n})$ is four-valued while $\omega(\vec{k})$ is multivalued but not always four-valued.

\begin{figure}[htb]
    \includegraphics[width=\linewidth]{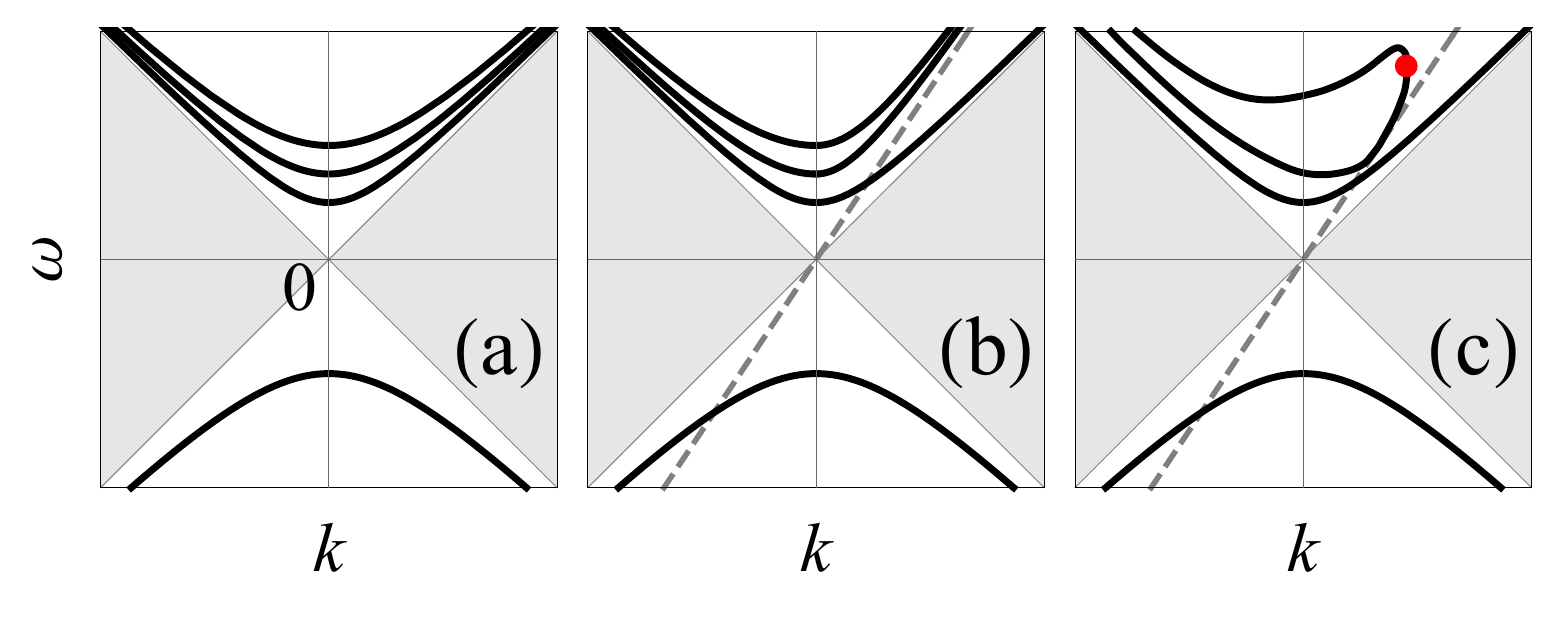}
    \caption{
        Schematic pictures of the DRs for the cases in which (a) $\omega$ is real for all $\vec{n}\in B$, (b) $\omega$ diverges at $\vec{n} \in B$ and (c) $\omega$ has a branch point (red dot).
        The black solid lines show $\omega(\vec{k} = k\vec{e}) \in \mathbb{R}$, where we choose some direction $\vec{e}$, and the gray regions are the zones of avoidance.
    }
    \label{fig:DRPattern}
\end{figure}

In the following discussions, we assume that $G(\vec{v})$ takes both positive and negative values and prove that $\omega(\vec{k})$ can be nonreal for some $\vec{k}\in\mathbb{R}^3$.
This proposition can be shown by proving the following 3 lemmas instead:
\begin{lem}
    If some of the 4 branches of $\omega(\vec{n} = \vec{0})$ are nonreal, there is a nonreal $\omega(\vec{k})$ for some $\vec{k} \in \mathbb{R}^3$.
\end{lem}
\begin{lem}
    $\omega(\vec{n})$ is nonreal for some $\vec{n}\in B$.
\end{lem}
\begin{lem}
    $\omega(\vec{n})$ does not diverge to infinity for all $\vec{n}\in B$.
\end{lem}
From lemma~1, we have only to consider the case in which all 4 branches of $\omega(\vec{n} = \vec{0})$ to be real; otherwise, the proposition is already proven.
Then, the DR with real $\omega$ can be categorized into 3 cases as Fig.~\ref{fig:DRPattern} by paying attention to $\vec{n} = \vec{k} / \omega$.
If all the branches of $\omega(\vec{n})$ are real for all $\vec{n} \in B$ [case~(a)] or some branches of $\omega(\vec{n})$ diverge to infinity for some $\vec{n}\in B$ [case~(b)], $\omega(\vec{k})$ is not necessarily nonreal;
otherwise, some branches of $\omega(\vec{n})$ must merge for some $\vec{n}\in B$ [case~(c)] and a branch point, at which the gradient $\nabla \omega(\vec{k})$ diverges and nonreal $\omega(\vec{k})$ begins, appears at some $\vec{k}$.
Since lemma~2 excludes case~(a) and lemma~3 excludes case~(b), the 3 lemmas lead to the existence of nonreal $\omega(\vec{k})$ for some $\vec{k}\in\mathbb{R}^3$, which is the proposition to prove.

Lemma~1 can be easily proven;
since $\vec{k} = \omega\vec{n}$ yields $\omega(\vec{k} = \vec{0}) = \omega(\vec{n} = \vec{0})$, the existence of nonreal $\omega(\vec{n} = \vec{0})$ immediately means that of nonreal $\omega(\vec{k} = \vec{0})$.
Lemma~3 is also confirmed from Eq.~(\ref{eq:Quartic}) because $\tr\mat T^m(\vec{n})$ is finite for $\vec{n} \in B$.
In the following, we prove lemma~2 by showing that there exists $\vec{n}\in B$ such that the coefficient of $\omega^2$ of $\tilde\Delta(\omega, \vec{n})$ is positive;
for such $\vec{n}$, $\tilde\Delta(\omega, \vec{n})$ has only 1 local minimum for $\omega$, meaning that the number of real solutions of the quartic equation $\tilde\Delta(\omega, \vec{n}) = 0$ is at most 2 and that the remaining solutions are nonreal.

\begin{figure*}[htb]
    \subfigure[]{
            \includegraphics[width=0.31\linewidth]{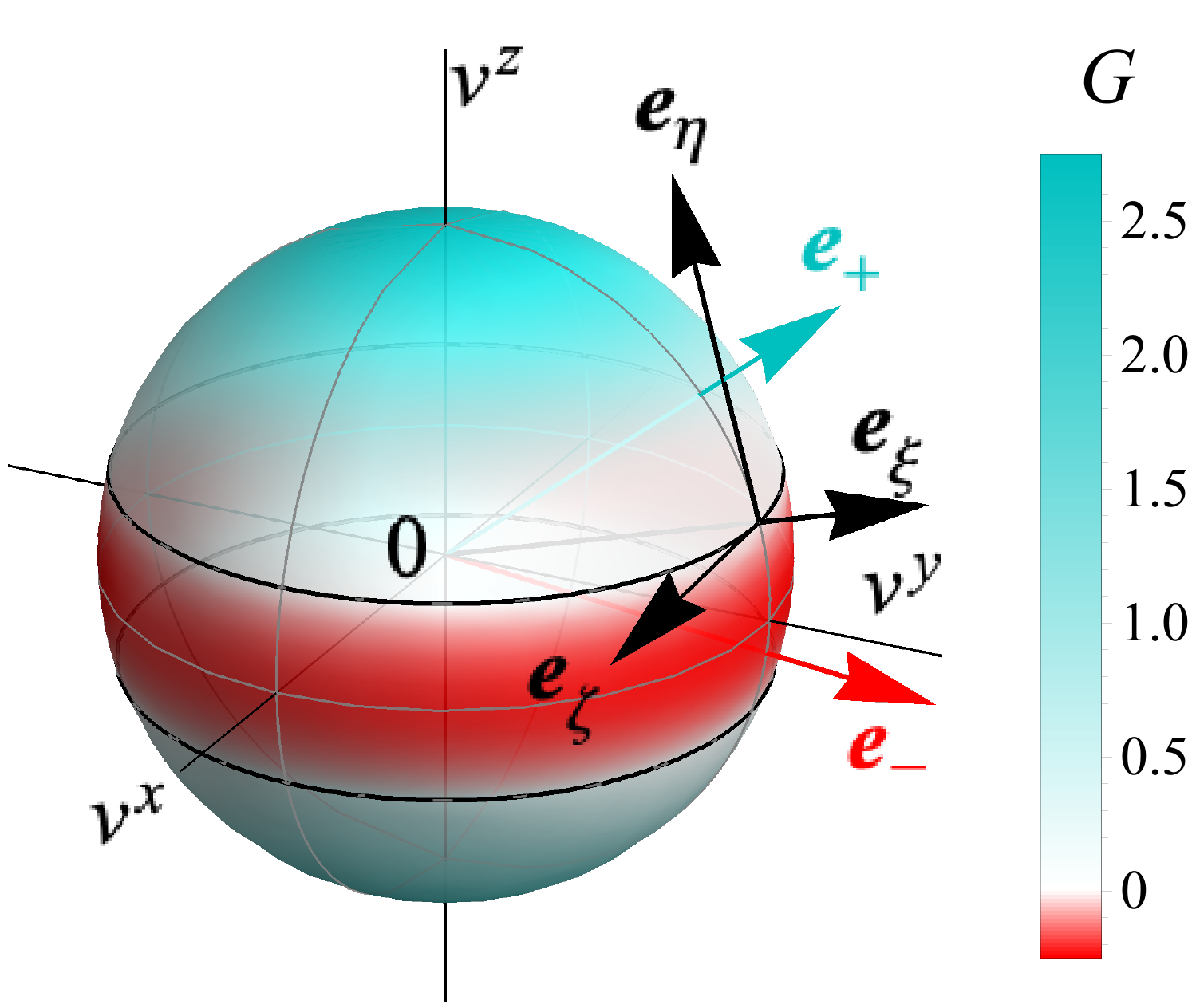}
            \label{fig:Distribution}
    }
    \subfigure[]{
            \includegraphics[width=0.31\linewidth]{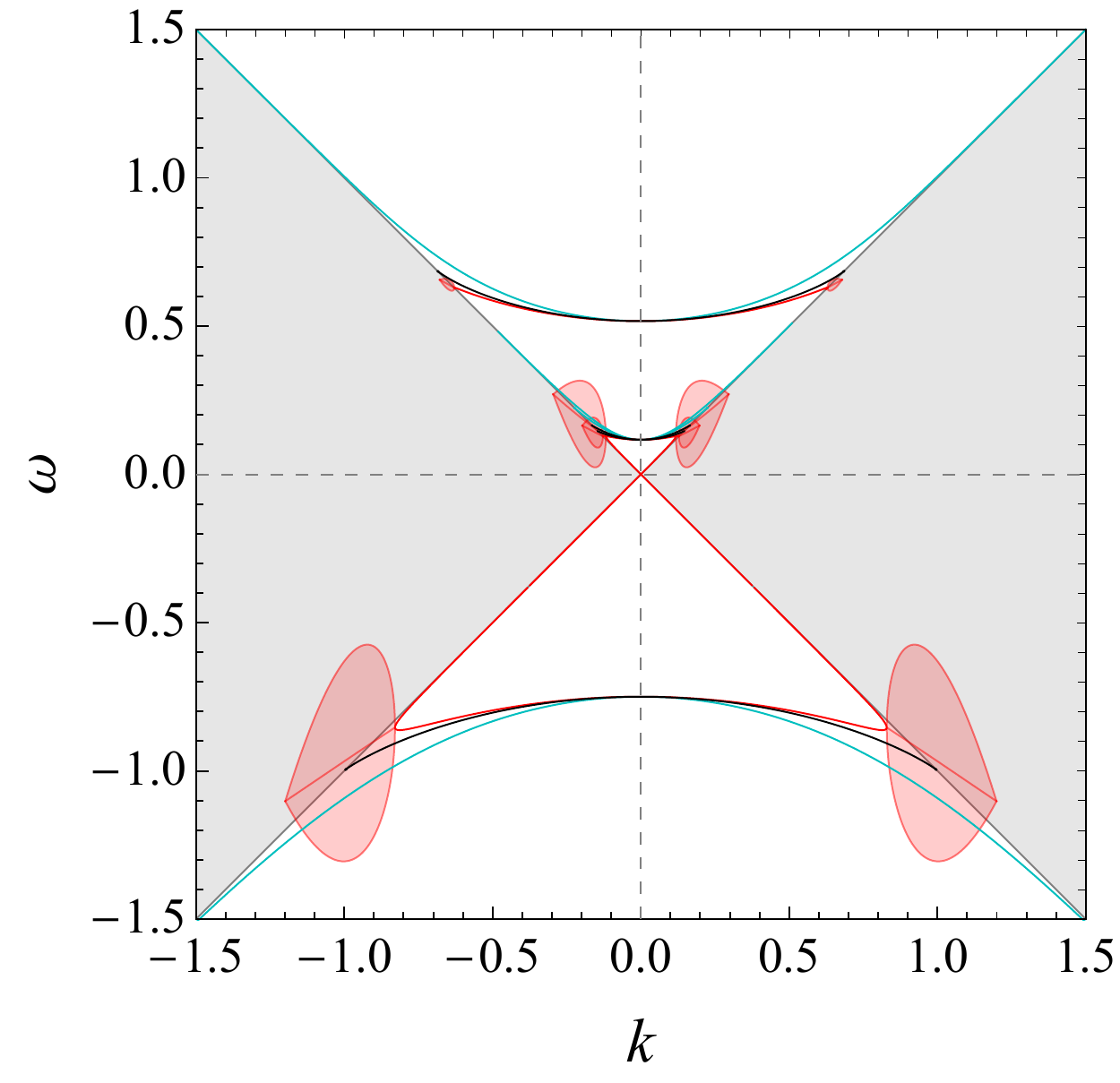}
            \label{fig:DR}
    }
    \subfigure[]{
            \includegraphics[width=0.31\linewidth]{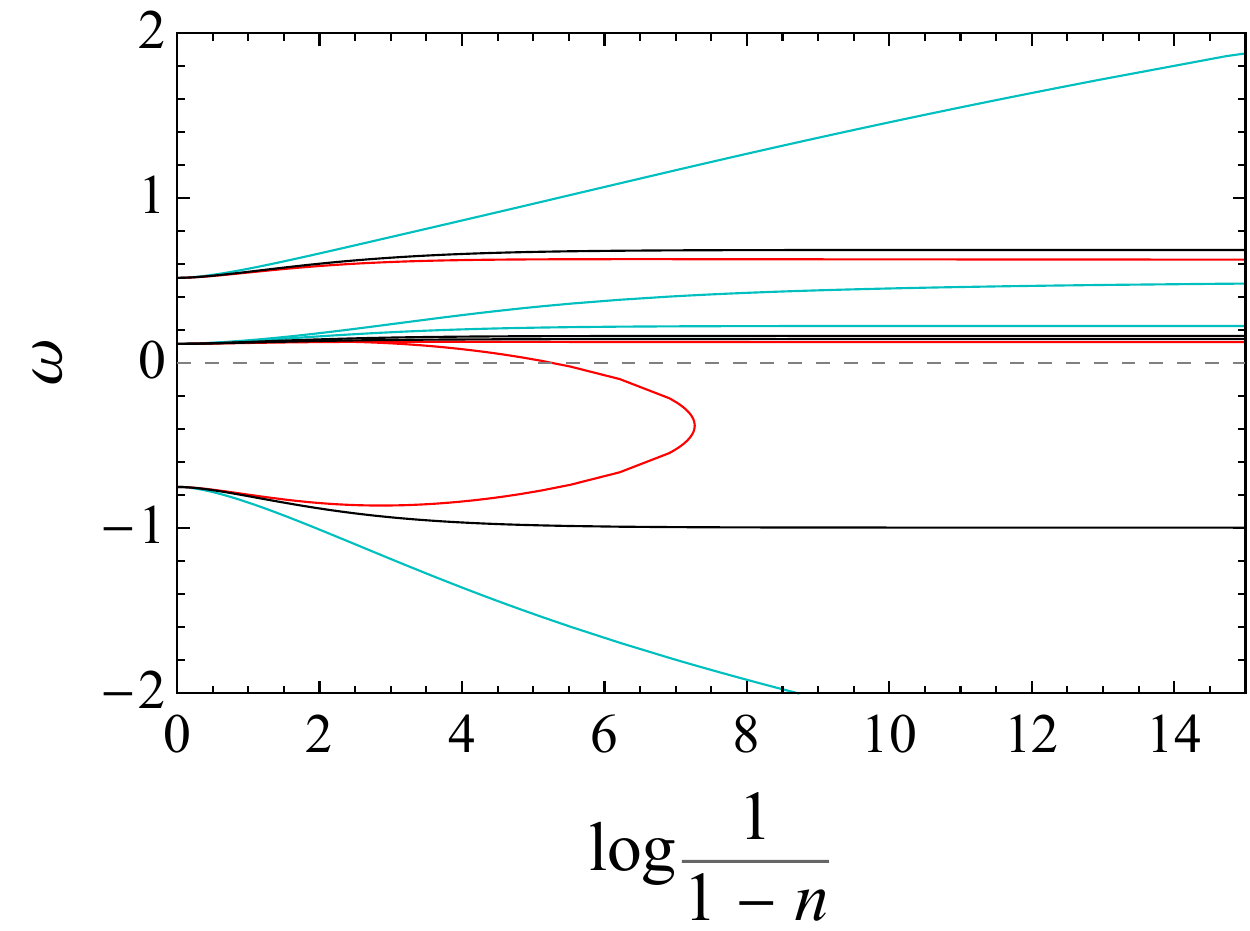}
            \label{fig:Asymptotics}
    }

    \caption{
        \subref{fig:Distribution}: The difference between the NFLN angular distributions of 2 flavors $G(\vec{v}) = 3(v^z)^2-\frac{1}{4}$ and $\{\vec{e}_\xi, \vec{e}_\eta, \vec{e}_\zeta\}$ and $\vec{e}_\pm$.
        The black solid lines on the sphere show the crossing directions.
        The scales of $\vec{e}_{\xi/\pm}$ are adjusted for visibility.
        \subref{fig:DR}: The DR for $G(\vec{v}) = 3(v^z)^2-\frac{1}{4}$.
        The black, cyan and red lines are $\omega(k\vec{e}_\xi)$, $\omega(k\vec{e}_+)$ and $\omega(k\vec{e}_-)$, respectively. The complex $\omega(k\vec{e}_-)$ values for real $k$ are indicated by the red areas, whose centerlines are $\re\omega$, and the difference between the centerlines and the boundaries of the areas is $10\im\omega$.
        The gray regions are the zone of avoidance.
        \subref{fig:Asymptotics}: The relation between $\omega$ and $\vec{n}$ for $G(\vec{v}) = 3(v^z)^2-\frac{1}{4}$.
        The directions of $\vec{n}$ are $\vec{e}_\xi$ (black), $\vec{e}_+$ (cyan) and $\vec{e}_-$ (red).
    }
	\label{fig:ToyModel}
\end{figure*}

We define $\vec{e}_\xi$ as one of the unit vectors satisfying $G(\vec{e}_\xi) = 0$;
we refer to these directions as crossing directions.
$\vec{e}_\eta$ is also defined as a unit vector parallel to $\vec{\nabla}G(\vec{e}_\xi)$, and $\vec{e}_\zeta \equiv \vec{e}_\xi \times \vec{e}_\eta$ (see Fig.~\ref{fig:Distribution}).
Hereinafter, the indices $t$, $\xi$, $\eta$ and $\zeta$ of vectors and tensors are used to denote their temporal, $\vec{e}_\xi$, $\vec{e}_\eta$ and $\vec{e}_\zeta$ components, respectively.

We focus on the behaviors of $\omega(\vec{n})$ around the crossing direction by considering
\begin{align}
    T^{\mu\nu}(n \mat{R}_\zeta(\theta)\vec{e}_\xi) =& \int \frac{d\vec{v}}{4\pi} G(\vec{v})\dfrac{v^\mu v^\nu}{1 - n\vec{v}\cdot\left\{\mat{R}_\zeta(\theta)\vec{e}_\xi\right\}}\nonumber\\
    =& R_\zeta{}^\mu{}_\sigma(\theta) R_\zeta{}^\nu{}_\rho(\theta)\tilde{T}_\theta^{\sigma\rho}(n),
\end{align}
where $\mat{R}_\zeta(\theta)$ is the rotation operator around $\vec{e}_\zeta$ with the angle $\theta$ and
\begin{align}
    \tilde{T}_\theta^{\mu\nu}(n) \equiv& \int \frac{d\vec{v}}{4\pi} G\left(\mat{R}_\zeta(\theta)\vec{v}\right)\dfrac{v^\mu v^\nu}{1 - n v^\xi}.
    \label{eq:TTilde}
\end{align}
Now, $\tr\mat{T}^m
(n \mat{R}_\zeta(\theta)\vec{e}_\xi) = \tr\mat{\tilde{T}}_\theta^m(n)$ is satisfied because $\mat{R}_\zeta$ is a Lorentz transformation.

$n > 1$ corresponds to the ``zone of avoidance"~\cite{Izaguirre2017}, in which $\mat{\tilde{T}}$ diverges to infinity and there is no solution satisfying Eq.~(\ref{eq:Quartic}).
At the limit of $n \uparrow 1$, $\mat{\tilde{T}}_\theta(n)$ seems to diverge as well.
All the components of $\mat{\tilde{T}}_\theta(n)$ whose indices include $\eta$ or $\zeta$, however, converge to a finite value because $v^\eta$ and $v^\zeta$ are proportional to $\sqrt{1 - \left(v^\xi\right)^2}$.
On the other hand, the other components diverge and asymptotically behave as
\begin{align}
    \tilde{T}_\theta^{tt}(n) \sim \tilde{T}_\theta^{t\xi}(n) \sim \tilde{T}_\theta^{\xi\xi}(n) \sim \frac{G\left(\mat{R}_\zeta(\theta)\vec{e}_\xi\right)}{2}\log\frac{1}{1 - n}
    \label{eq:AsymptoticBehavior}
\end{align}
as $n \uparrow 1$ for $\theta = \pm \epsilon$ with small $\epsilon > 0$.
At $\theta = 0$, $v^\xi = 1$ is zero for $G$, and all the components of $\mat{\tilde{T}}_\theta(n)$ converge to a finite value as $n \uparrow 1$.

The components that converge at $\theta = \pm\epsilon$ as $n \uparrow 1$ are continuous for $\theta$ at $\theta = 0$ and $n = 1$.
From Eq.~(\ref{eq:AsymptoticBehavior}), the other components for $\theta = \epsilon$ and $\theta = - \epsilon$ diverge to infinity with different signs from each other as $n \uparrow 1$.
Although these components diverge, the differences
\begin{align}
    c_\theta(n) \equiv& \tilde{T}_\theta^{t\xi}(n) - \tilde{T}_\theta^{tt}(n) = - \int \frac{d\vec{v}}{4\pi} G\left(\mat{R}_\zeta(\theta)\vec{v}\right)\dfrac{1 - v^\xi}{1 - n v^\xi}\\
    d_\theta(n) \equiv& \tilde{T}_\theta^{\xi\xi}(n) - \tilde{T}_\theta^{tt}(n) = - \int \frac{d\vec{v}}{4\pi} G\left(\mat{R}_\zeta(\theta)\vec{v}\right)\dfrac{1 - (v^\xi)^2}{1 - n v^\xi}
\end{align}
converge to a finite value as $n \uparrow 1$, and hence, those as $\theta \uparrow 0$ and $\theta \downarrow 0$ coincide with each other at $n = 1$.
Then, straightforward computation yields asymptotic behavior
\begin{align}
    -\frac{1}{2}\tr\mat{\tilde{T}}_\theta^2(n) \sim \left[2c_\theta(1) - d_\theta(1)\right]\tilde{T}_\theta^{tt}(n)
\end{align}
as $n \uparrow 1$.
This is the coefficient of $\omega^2$ in Eq.~(\ref{eq:Quartic}) and takes positive values for either $\theta = \epsilon$ or $\theta = -\epsilon$ with sufficiently large $n$~\footnote{
    If $d_0(1) - 2c_0(1)$ vanishes for all crossing directions, we have to consider subleading terms and/or other coefficients in Eq.~(\ref{eq:Quartic}).
    We do not go into further detail because such a case is quite special with measure zero.
}.
Therefore, for at least one of $\theta = \epsilon$ or $\theta = -\epsilon$, there exist nonreal $\omega$ values for sufficiently large $n$, and lemma 2 has been proven.

We note that the proof of the sufficient condition here is not valid for discrete spectra, unlike the case of the necessary condition.
Whether the sufficient condition holds also for the discrete case is left for future research.

We focus on the distribution $G(\vec{v}) = 3(v^z)^2 - \frac{1}{4}$ to exemplify the above discussion (see Fig.~\ref{fig:Distribution}).
In this case, all the points satisfying $v^z = \pm\frac{1}{2\sqrt{3}}$ are the crossing directions.
Here, we choose $\{\vec{e}_\xi, \vec{e}_\eta, \vec{e}_\zeta\}$ as
\begin{align}
    \begin{pmatrix}
        \vec{e}_\xi\\
        \vec{e}_\eta\\
        \vec{e}_\zeta
    \end{pmatrix}
    =
    \begin{pmatrix}
        0 & \frac{\sqrt{11}}{2\sqrt{3}} & \frac{1}{2\sqrt{3}}\\
        0 & -\frac{1}{2\sqrt{3}} & \frac{\sqrt{11}}{2\sqrt{3}}\\
        1 & 0 & 0
    \end{pmatrix}
    \begin{pmatrix}
        \vec{e}_x\\
        \vec{e}_y\\
        \vec{e}_z
    \end{pmatrix}
\end{align}
and define $\vec{e}_{\pm} \equiv \mat{R}_\zeta(\pm \pi/8)\vec{e}_\xi$.

The DRs for $\vec{k}$ parallel to $\vec{e}_{\xi/\pm}$ are shown in Fig.~\ref{fig:DR}.
We can confirm that nonreal $\omega$ values appear only for $\vec{k} = k\vec{e}_-$ and begin at the points at which $d\omega / dk$ diverges to infinity.
We note that the solution $\omega(\vec{k})$ can vanish at large $|\vec{k}|$.
If $\Delta(k)$ was holomorphic on $\mathbb{C}^4$, $\omega(\vec{k}) \in \mathbb{C}$ would exist for all $\vec{k}\in\mathbb{R}^3$.
In reality, however, $\Delta(k)$ has the branch cut on $\omega \in (-|\vec{k}|, |\vec{k}|)$, which corresponds to the zone of avoidance, and the zeros of $\Delta(k)$ can terminate on the branch cut.

Figure~\ref{fig:Asymptotics} shows the asymptotic behaviors of $\omega$ as $n \uparrow 1$ for $\vec{n}$ parallel to $\vec{e}_{\xi/\pm}$.
For $\vec{n} = n\vec{e}_\xi$, all the branches of $\omega(\vec{n})$ converge to finite values as $n \uparrow 1$ because all the components of $\mat{T}$ converge.
On the other hand, for $\vec{n} = n\vec{e}_+$, only 2 of them converge, and the remaining 2 logarithmically diverge to infinity because some components of $\mat{T}$ diverge, as shown in Eq.~(\ref{eq:AsymptoticBehavior}).
For $\vec{n} = n\vec{e}_-$, while 2 branches converge, the remaining 2 merge at $n \approx 1 - e^{-7.3}$, and $\omega$ become nonreal for $n$ larger than this branch point;
for sufficiently large $n$, the number of real branches is less than 4, which is the number of real branches of the $\vec{k} = \vec{0}$ mode, meaning that there is some branch point of $\omega(k\vec{e}_-)$ for $k \in \mathbb{R}$.

\section{Conclusion}
We showed that fast flavor instability is present if and only if the NFLN angular distributions of 2 flavors cross each other.
To find fast instability, we have only to seek NFLN crossings.
In contrast, once an NFLN crossing appears, the flavor coherence grows in the linear regime, and nonlinear oscillations are expected to begin after several times the linear growth timescale.

We also find that unstable modes appear at least in $\vec{k}$ around the crossing directions.
We have to consider that the fast instabilities may not be able to be captured if some symmetries are imposed a priori.
Determining which modes are actually unstable is important for reasonable results when we conduct nonlinear calculations.

To crystallize the effect of collective neutrino oscillations on astrophysical systems, nonlinear behaviors should also be elucidated.
The resultant distributions after a sufficiently long time in the regions where instabilities have propagated might be simply flavor-decohered distributions.
Whatever the results of nonlinear evolutions are, it is important to accurately understand the behaviors in the linear regime, including how instabilities propagate in spacetime~\cite{Sturrock1958,Briggs1964,Landau1997,Capozzi2017,Yi2019,Morinaga2020,Morinaga2021}, and this study has achieved one of the major goals toward this understanding.

\begin{acknowledgments}
I am grateful to Shoichi Yamada for his valuable comments.
I would also like to thank Georg Raffelt, Basudeb Dasgupta, Sajad Abbar, and Manu George for their useful discussions and comments.
I am supported by a JSPS Grant-in-Aid for JSPS Fellows (No. 19J21244) from the Ministry of Education, Culture, Sports, Science and Technology (MEXT), Japan.
\end{acknowledgments}

\appendix
\section{Proof of equation~(\ref{eq:Quartic})}
We consider the characteristic polynomial $p_{\mat{A}}(z) \equiv \det(z \mat{I}_N - \mat{A})$ for $N\times N$ matrix $\mat{A}$.
$\tilde\Delta(\omega, \vec{n})$ in Eq.~(\ref{eq:Quartic}) is $\tilde\Delta(\omega, \vec{n}) = p_{-(T^\mu{}_\nu)(\vec{n})}(\omega)$.
In general, the coefficients of $p_{\mat{A}}(z)$ can be expressed as the summation of products of $\tr\mat{A}^k\ (k = 1, \cdots, N)$, and their explicit expressions can be calculated by the recursion formula derived below~\cite{Silva1998}.

$p_{\mat{A}}(z)$ is factorized as $p_{\mat{A}}(z) = \prod_{i = 1}^{N}(z - \lambda_i)$, where $\lambda_1, \cdots, \lambda_N$ are the eigenvalues of $\mat{A}$.
It is also expanded as $p_{\mat{A}}(z) = \sum_{i = 0}^{N}\tilde{e}_i(\lambda_1, \cdots, \lambda_N) z^{N - i}$, where $e_k(x_1, \cdots, x_n)$ is the elementary symmetric polynomial of degree $k$ in $n$ variables $x_1, \cdots, x_n$ and $\tilde{e}_k \equiv (-1)^k e_k$.

The elementary symmetric polynomials satisfy Newton's identity
\begin{align}
    \sum_{i = 0}^{k - 1}&\, p_{k - i}(x_1, \cdots, x_n) \tilde{e}_i(x_1, \cdots, x_n)\nonumber\\
    &+ k \tilde{e}_k(x_1, \cdots, x_n) = 0\quad \mathrm{for}\ k = 1, \cdots, n,
\end{align}
where $p_k(x_1, \cdots, x_n) \equiv \sum_{i = 1}^n x_i^k$.
Since $p_k(\lambda_1, \cdots, \lambda_N) = \tr \mat{A}^k$ is satisfied, $\tilde{e}_k(\lambda_1, \cdots, \lambda_N)$, which is the coefficients of $p_{\mat{A}}(z)$, is given by
\begin{align}
    \tilde{e}_k(\lambda_1, \cdots, \lambda_N) = -\frac{1}{k}\sum_{i = 0}^{k - 1} \tr\mat{A}^{k - i}\tilde{e}_i(\lambda_1, \cdots, \lambda_N)\nonumber\\
    \mathrm{for}\ k = 1, \cdots, N.
\end{align}
Beginning with $\tilde{e}_0(\lambda_1, \cdots, \lambda_N) = 1$, $\tilde{e}_k(\lambda_1, \cdots, \lambda_N)$ for $k = 1, \cdots, N$ can be calculated recursively.

\section{Proof of equation~(\ref{eq:AsymptoticBehavior})}
To prove Eq.~(\ref{eq:AsymptoticBehavior}), we consider
\begin{align}
    I(n) \equiv& \int \frac{d\vec{v}}{4\pi} \frac{f(\vec{v})}{1 - n v^\xi},
\end{align}
where $f$ is a continuous function on the unit sphere that satisfies $f(\vec{e}_\xi) \neq 0$.
This integral can be decomposed as
\begin{align}
    I(n) =& \int_{-1}^1\frac{dv^\xi}{2}\int_0^{2\pi}\frac{d\phi}{2\pi}\dfrac{f(\vec{v})}{1 - n v^\xi}\nonumber\\
    =& \int_{-1}^1\frac{dv^\xi}{2}\frac{F(v^\xi)}{1 - n v^\xi},
\end{align}
where $\phi$ is the azimuthal angle when we choose the zenith as the $\vec{e}_\xi$-direction and $F(v^\xi) \equiv \int_0^{2\pi}\frac{d\phi}{2\pi}f(\vec{v})$.
For arbitrary $n \in (-1, 1)$, by dividing the integral domain at $v^\xi = 1 - \epsilon$ with $\epsilon > 0$ and applying the mean-value theorem, we obtain $v^\xi_- \in [-1, 1 - \epsilon]$ and $v^\xi_+ \in [1 - \epsilon, 1]$ such that $I$ is expressed as
\begin{align}
    I(n) =& F(v^\xi_{-})\frac{1}{2n}\log\frac{1 + n}{1 - n(1 - \epsilon)}\nonumber\\
    &+ F(v^\xi_+)\frac{1}{2n}\log\frac{1 - n(1 - \epsilon)}{1 - n}.
\end{align}
Regardless of how small of $\epsilon$ we choose, the second term dominates at the limit of $n \uparrow 1$.
Because $F(v^\xi_+) \to F(1) = f(\vec{e}_\xi)$ as $\epsilon \downarrow 0$, asymptotic behavior
\begin{align}
    I(n) \sim \frac{f(\vec{e}_\xi)}{2}\log\frac{1}{1 - n}
\end{align}
is obtained.


%

\end{document}